\documentclass[twocolumn]{article}

\usepackage[colorinlistoftodos]{todonotes}
\usepackage{authblk}
\usepackage{hyperref}
\usepackage{balance}
\title{Discussion of Intelligent Electric Wheelchairs for Caregivers and Care Recipients\\}

\author{Satoshi Hashizume}
\author{Ippei Suzuki}
\author{Kazuki Takazawa}
\author{Yoichi Ochiai\thanks{wizard@slis.tsukuba.ac.jp}}
\affil{University of Tsukuba}

\date{}

\begin{document}
\maketitle

\begin{abstract}
In order to reduce the burden on caregivers, we developed an intelligent electric wheelchair. 
We held workshops with caregivers, asked then regarding the problems in caregiving, and developed problem-solving methods. 
In the workshop, caregivers' physical fitness and psychology of the older adults were found to be problems and a solution was proposed. 
We implemented a cooperative operation function for multiple electric wheelchairs based on the workshop and demonstrated it at a nursing home. 
By listening to older adults, we obtained feedback on the automatic driving electric wheelchair. 
From the results of this study, we discovered the issues and solutions to be applied to the intelligent electric wheelchair.
\end{abstract}


\section{Introduction}
Recently, research and development work has examined practical applications of automotive driving technology. 
Automotive driving requires recognition of the environment, judgment of the situation, and operations planning by a machine instead of a human driver. 
Therefore, advanced information processing technologies such as artificial intelligence (AI) and sensor fusion are required. 
Technology in automatic driving can also be applied to machines other than cars, such as factory machines and personal mobility. 
However, with current technology, it is still difficult to fully automate the drive function for personal mobility machines; hence, they have not been put to practical use on public roads.

Wheelchairs are important modes of mobility for older adults and for physically handicapped persons with limited movement. 
Electric wheelchairs began appearing on the market in the 1950s. 
Since then, people who have difficulty driving non-electric wheelchairs, due to lack of physical strength can easily drive electric wheelchairs. 
In nursing care facilities, it is necessary for the caregiver to remain beside the electric wheelchair at all times. 
The global population aged 60 years or over numbered 962 million in 2017. 
The number of older persons is expected to double again by 2050, when it is projected to reach nearly 2.1 billion \cite{WPA2017}.
In these aging societies, there are increasingly urgent demands for wheelchairs and caregivers. 
In addition, because of a shortage of caregivers, the burden on them is also increasing. 
How to reduce the burden on caregivers is a challenging problem.

\begin{table*}[t]
\caption{Overview of our experiments}
\vspace{1em}
\centering
\begin{tabular}{l|c|c|l}
     & \begin{tabular}{l}
     Number of\\participants
     \end{tabular}& Target & Purpose \\ \hline
    \begin{tabular}{l}
     St1-\\Problem finding
     \end{tabular}
     & 6 & Caregivers &  \begin{tabular}{l}To find the problems and solutions in care.\end{tabular} \\ \hline
    \begin{tabular}{l}
     St1-\\Problem solving
     \end{tabular} & 2 & Caregivers & \begin{tabular}{l}
     To evaluate methods found\\during the problem finding phase.
     \end{tabular}\\ \hline
    \begin{tabular}{l}
     St2
     \end{tabular}
     & 2 & \begin{tabular}{l}
     Care\\recipients
     \end{tabular} & \begin{tabular}{l}
     Verbal survey of older adult people\\ who experience the intelligent wheelchair.
     \end{tabular}
  \end{tabular}
  \label{tab:exp_overview}
\end{table*}

To reduce this burden, a fully-automated electric wheelchair is ideal for caregivers who use wheelchairs in their work. 
With a fully-automatic wheelchair, a caregiver does not always have to accompany the care recipient. 
Furthermore, the care recipient does not have to request assistance; thus, the automatic wheelchair reduces their psychological burden. 
Currently, however, automatic driving techniques are not widely adopted in nursing care fields.

Therefore, we developed an intelligent electric wheelchair called the Telewheelchair. 
In previous research, we conducted an operability experiment evaluation, where the usefulness of the remote control was demonstrated. 
Because the Telewheelchair was developed with general-purpose equipment and software, it is possible to add additional functions. 
In the experiment, We also discussed what new functions should be added to solve problems, specifically in the nursing care environment.

To identify the functions to be installed in an intelligent electric wheelchair, we explored the types of problems encountered in a practical nursing care environment. 
We conducted two-part studies (Table \ref{tab:exp_overview}) at a nursing home using the intelligent wheelchair. 
The first study aimed to find the problems and their solutions in the nursing care facility. 
Based on the results of that study, we implemented the cooperative operation function with the intelligent wheelchair and demonstrated it at the nursing home. 
The second study was a verbal survey of older adults who experienced the intelligent wheelchair.

Contributions of this research are as follows.
\begin{itemize}
\item We held a workshop at a nursing home, and from the feedback of the experiments, we found that there was a problem of the workload of caregivers in nursing care.
\item We proposed a cooperative operation. Based on the implementation and feedback from the experiments, we concluded it was a useful method to solve the problems in nursing care.
\item We demonstrated the remote control for the older adults and found that they have positive opinions of the intelligent wheelchair.
\end{itemize}

\section{Related Work}
\subsection{Technical approach for an aging society}
In an aging society, psychological and physical burdens are placed on the older adults and caregivers. 
Many technological implementations are under way to reduce the burden on these groups.
The portal monitor \cite{Caldeira:2017:SCA:2998181.2998206} is a monitoring system that addresses the privacy of older adults.
Safety and consideration of the psychological problems of older adults have been studied.
Ichinotani et al. \cite{Ichinotani:2018:MCT:3192975.3192983} developed a deformable wheelchair and a nursing care bed. 
The systems are used to solve the problem of older adults getting hurt when transferring  from a bed to a wheelchair. 
Huang et al. \cite{Huang:2015:VLR:2829966.2829971} conducted an approach to solve the loneliness and anxiety of the older adult through talking with CG animation agents. 
This method showed that active listening could be done at the same level as human beings. 
Conte et al. \cite{Conte:2018:ITA:3170427.3186548} proposed a system that provided  tactile assistance for a tablet according to the learning preferences of older adults.

In addition, research has been conducted to investigate the relationship between technology and the aging society. 
Senior Care for Aging in Place \cite{Caldeira:2017:SCA:2998181.2998206} investigated how older adults live with both self-care and collateral care at the same time. 
They found that many older adults need a monitoring system that does not violate their privacy. 
Technological Caregiving \cite{Piper:2016:TCS:2858036.2858260} conducted an interview survey on the caregiver side and the nursing care side about online support for older adults.
They showed that online nursing care activities are generally stressful to caregivers.
Buccoliero et al. \cite{Buccoliero:2014:ASE:2691195.2691303} aimed to clarify the factors that affect the adoption of technology concerning the health of older adults. 
Kostoska et al.~\cite{Kostoska:2018:RSA:3227609.3227686} aimed to replace manual data collection with automatic data collection using a distributed system based on mobile phones and biosensors. 
Researchers, acting as proxies for informal caregivers, \cite{Davis:2014:RPI:2686612.2686652} examined the specifications of PictureFrame, a new social technology to support the provision of medical examinations for older adults residing at home. 
Virtual Carer \cite{Moreno:2016:VCP:2910674.2935855}, a web service model, generated personalized recommendations for informal caregivers according to the need of older adults. 
Arif et al. \cite{Arif:2017:CES:3077584.3077598} developed a mobile telepresence robot, ``Telemedical Assistant'', that minimized medical mistakes.
Chen et al. \cite{Chen:2013:CCD:2441776.2441789} conducted an interview study on caregivers. 
Caregivers maintained the balance of their personal lives with work, family, and their caregiver roles with the concept of giving-impact and visibility-invisibility. 
Holbo et al. \cite{Holbo:2013:SWT:2513383.2513434} examined factors to allow people with dementia to walk safely from workshops and interview.

\begin{figure}[tb]
	\centering
    \includegraphics[width=\linewidth]{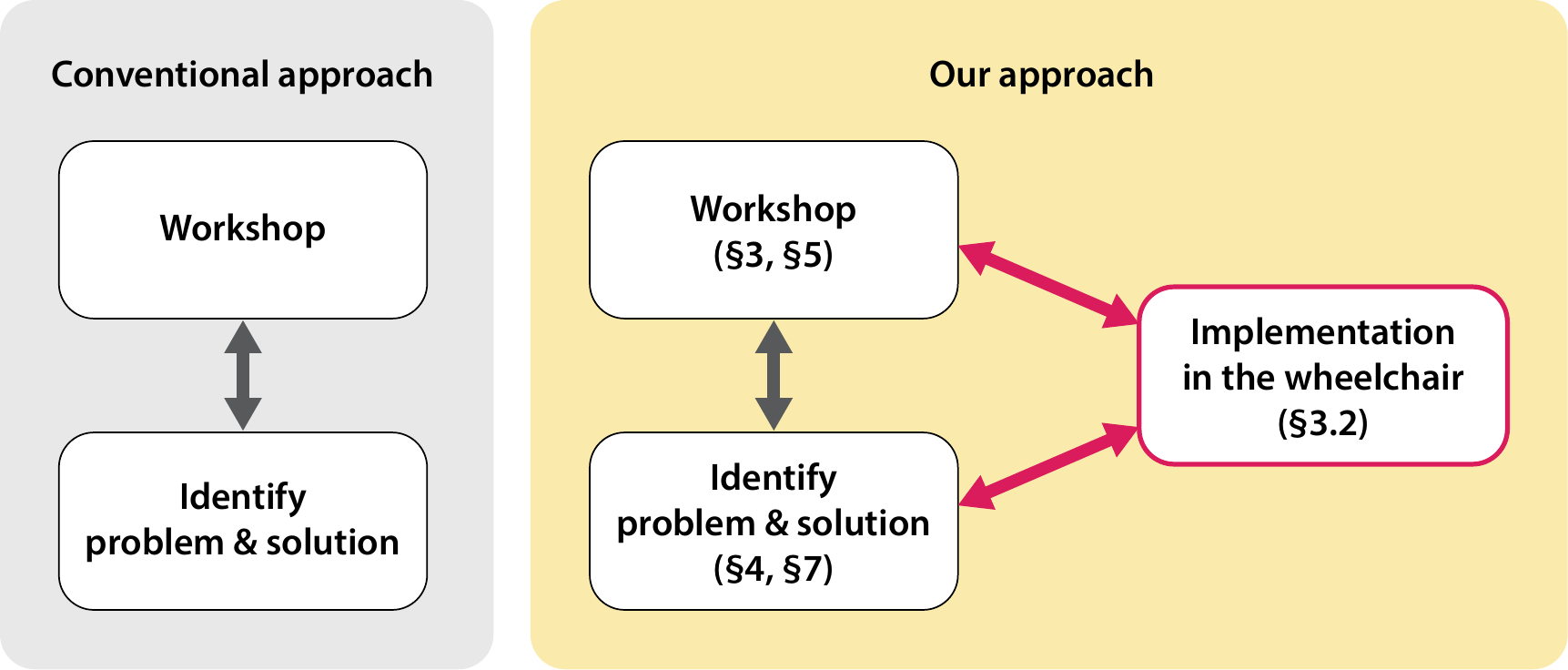}
    \caption{Position of our study. Our approach used turning loop of problem definition, implementation, and workshop.}
    \label{fig:position}
\end{figure}

Many technologies for the aging society were implemented as described above, and the relationship between older adults, caregivers, and technical approaches has been investigated from various perspective. 
However, most of these studies had been limited to finding problems in care from workshops and did not implement functions. 
We implemented one of the solutions from the proposed problem and held a workshop again. 
We can solve the problem quickly by turning loop of problem definition, implementation, and workshop (Figure \ref{fig:position}).
Moreover, there is little practical discussion of how technologies like automatic wheelchairs are perceived by older adult participants in nursing homes. 
We tried the approach of simulating the automatic driving concept in practical experiments to discover methods to reduce the burden on older adult people and caregivers.

\subsection{Intelligent electric wheelchair}
Since the 1990s, many researchers have been studying the automatic operation of electric wheelchairs. 
Gundersen et al. \cite{gundersen1996applications} presented an electric wheelchair equipped with a remote-control system via a head mounted display (HMD) and an obstacle detection system with an ultrasonic sensor. 
Pires et al. \cite{bento2002behavior} explored the usability of wheelchairs by conducting experiments on their operation using voice and joystick, and on obstacle detection and collision detection. 
The NavChair \cite{levine1999navchair} enhanced navigational functionality by enabling guided door passage and wall following in addition to obstacle avoidance. 
Mazo \cite{mazo2001integral} examined automatic driving by using various methods for environmental recognition and user motion detection.
Diao et al. \cite{8243508} developed the Intelligent Wheelchair Bed, which could detect and avoid obstacles using eleven ultrasonic sensors. 
Hua et al. \cite{HUA201714} developed a wheelchair that recognized the environment by analyzing laser range finder and camera data with a neural network.
Faria et al. \cite{Faria2015} developed a new wheelchair control method for Cerebral Palsy. 
Telewheelchair \cite{Hashizume:2018:TRC:3174910.3174914}, an electric wheelchair equipped with a remote control function, a computer-operation support function, and controlled by a HMD, was developed to reduce the burden on the caregivers. 

Some research focuses on designing a general-purpose electric wheelchair as well as automatic operation of specific motions. 
The DECoReS system~\cite{Hasegawa:2015:DDE:2814940.2814942} can be driven using orders such as ``go straight, fast'' and ``make a wide curve to the right'', turning corners according to the user's preference. 
Kobayashi et al. \cite{kobayashi2013robotic} uses laser range sensors to move the electric wheelchair while keeping a certain distance from the accompanying caregiver. 
Passengers can communicate with their companions.

\begin{figure*}[tb]
	\centering
    \includegraphics[width=\linewidth]{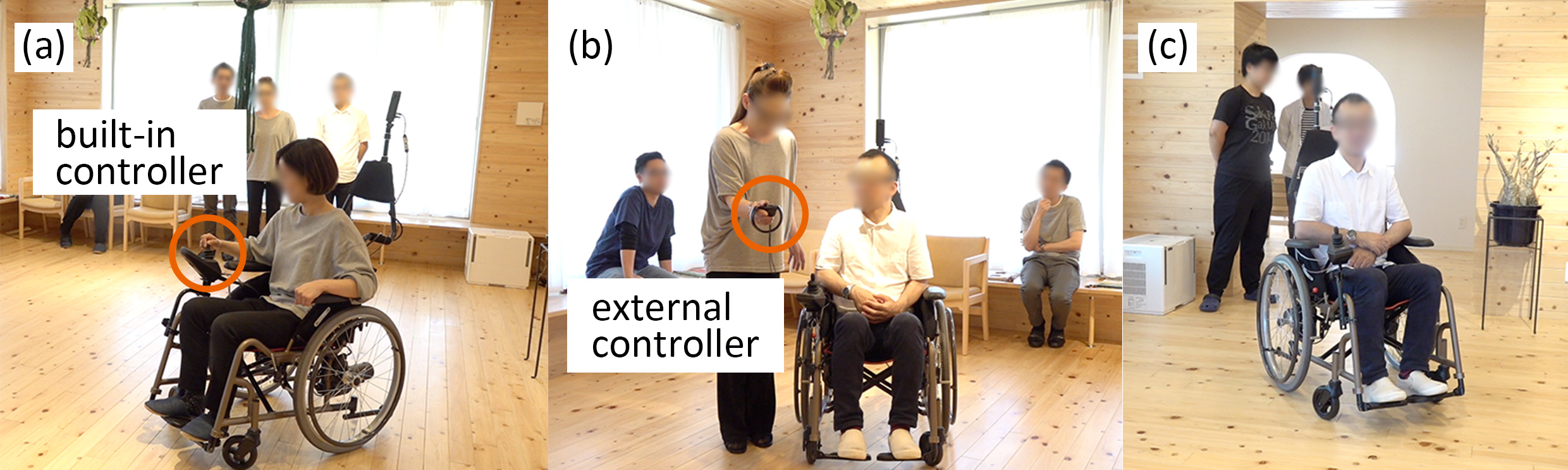}
    \caption{Demonstration of the following three types of electric wheelchair to participants: (a) participant rides on an electric wheelchair and operates it with a controller in hand, (b) participant is standing by the wheelchair and using the controller, and (c) participant rides on a wheelchair operated remotely.}
    \label{fig:demothreetype}
\end{figure*}

\section{Study 1: Problem Finding and Solving}

The purpose of this research was to discuss how the intelligent electric wheelchair can be utilized in the nursing care field. 
First, there was a need to know what problems are present in the nursing care field. 
Therefore, to identify these problems, we held a workshop at a nursing home. 
Next, we considered methods that could solve the problems. 
Lastly, we conducted experiments with an intelligent electric wheelchair. 
We proposed to discuss the intelligent electric wheelchair more practically by conducting a series of steps from problem finding to practical implementation.

\subsection{Problems finding phase}
\subsubsection{Design}

The purpose of this research was to reduce the burden on caregivers using an intelligent wheelchair. 
We needed to find out what kind of problems there are in the work of nursing care facilities. 
In the problem finding phase, we aimed to find the problems with wheelchair care at nursing care facilities. 
Participants of the workshop were staff working at nursing care facilities. 
Caregivers have a lot of experience and knowledge about nursing care, so it was possible to describe the correct problems in nursing care sites.
At the same time, subjects considered problem solving methods based on their experience at nursing care facilities. 
In problem solving phase, we implemented function into wheelchair based on problem solving method.

Caregivers participating in the workshop had little knowledge about automatic driving and artificial intelligence. 
Therefore, we also conducted a technical lecture on these topics during the workshop. 
Technical training included explanations on artificial intelligence and wheelchair operation, as well as an intelligent wheelchair demonstration. 
In the intelligent wheelchair demonstration, the caregivers participated in three types of experiments: (i) An occupant manipulated a joystick mounted on an electric wheelchair (Figure \ref{fig:demothreetype} (a)), (ii) A caregiver stood at a distance and controlled the wheelchair using the remote controller (Figure \ref{fig:demothreetype} (b)), and (iii) An operator simulated automatic driving (Figure \ref{fig:demothreetype} (c)). 
Experiments (i) and (ii) were based on the implementation of previous research \cite{Hashizume:2018:TRC:3174910.3174914}. 
Experiment (iii) was a mode in which a wheelchair automatically moved when the pilot is not nearby. 
It is noted that instead of implementing a self-driving electric wheelchair, we actually conducted an experiment that simulated the self-driving environment. 
An operator controlled the wheelchair from a remote location. 
However, the participants were not told that the control operation was manually controlled  from a remote location; therefore, participants believed that the electric wheelchair was being driven automatically.
The aim of experiment (iii) was to encourage participants to experience operations like automatic driving and to have deeper understanding of automatic driving.
The target comprehension level of participants after receiving technical training was an understanding of the meaning and purpose of artificial intelligence and the outline of the intelligent wheelchair. 
Participants were asked to consider solutions to problems before and after technical training. Following their technical training, we asked them to respond using the artificial intelligence information they learned in the course and their knowledge of automatic driving.

\subsubsection{Procedure}  
Each participant was briefly informed of the purpose of the study and was informed that they could stop the study and take a break at any time. 
Further, they were provided with a consent form to sign and a demographics questionnaire to complete.
At the beginning and end of the workshop, the participants were asked to answer their expectation level for AI and robot care using a five-level Likert scale. 

Initially, the participants were asked to write out their career experience and problems they have encountered.
The response time was 15 min. 
Next, the participants were asked to write a method capable of solving a nursing care problem. 
We asked them to write three problems that they wanted to solve, and their proposed solutions. 
The problems they wanted to solve were expected to be based on the problem that they wrote. 
The response time was 30 min. Next, we presented a technical training demonstration using the electric wheelchair. 
The duration of the technical course was 15 min and the demonstration was 45 min. 
After the three types of demonstrations were completed, participants were asked to provide their impressions of each demonstration. 
Lastly, we asked the participants to write a method capable of solving a nursing care problem. 
The second problem solving method was to be based on the material learned in the technical course. 
It included up to three problems that participants wanted to solve. 
The response time was 30 min. 
All response times were sufficient to answer the questions.

\subsubsection{Participants}
Six participants (three females and three males) with ages between 23 and 63 years (M = 39, SD = 14.2) participated in the experiment (Table \ref{tab:summary_caregivers}). 
All participants were caregivers. 
All participants were employed in nursing care facilities that conducted workshops. 
The average years of employment was 11.5 years (SD = 7.2). 
Four participants did not know about Telewheelchair in advance and two participants knew a little about it. 
All participants spoke Japanese.

\begin{table}[t]
\caption{Summary of workshop participants on intelligent wheelchair for caregivers. *``Chief'' is responsible for supervising caregivers in nursing homes}
\vspace{1em}
\centering
 \begin{tabular}{l|c|c|l|c}
     & Sex & Age & 
     Position
      & 
      \begin{tabular}{l}
      Length of \\service (year)
      \end{tabular}
      \\ \hline
    $P_{1}1$ & F & 29 & \begin{tabular}{l}Caregiver\end{tabular} & 7 \\ \hline
    $P_{1}2$ & F & 23 & \begin{tabular}{l}Former\\ caregiver\end{tabular} & 3 \\ \hline
    $P_{1}3$ & M & 33 & \begin{tabular}{l}Chief$^*$\end{tabular} & 4 \\ \hline
    $P_{1}4$ & M & 42 & \begin{tabular}{l}Chief\end{tabular} & 20 \\ \hline
    $P_{1}5$ & F & 63 & \begin{tabular}{l}Chief\end{tabular} & 20 \\ \hline
    $P_{1}6$ & M & 44 & \begin{tabular}{l}Chief\end{tabular} & 12 
  \end{tabular}
  \label{tab:summary_caregivers}
\end{table}

\subsection{Problems solving phase}
\subsubsection{Design}
 In this phase, the aim was to implement and evaluate methods found during the problem finding phase. 
Caregivers who participated in the problem problem finding phase raised a common problem: having fewer hours to care for older adults. 
For example, the duties of a caregiver consisted of a large number of tasks, such as working alone to assist three older adults with their diets, and assisting with the bathing of more than 30 people a day. 
With this workload, the caregiver may not had time to communicate well with the older adults. 
Hence, to increase the time spent for meals, nursing care, and bathing assistance, we considered shortening the traveling time of the wheelchair.

To shorten the travel time of the wheelchairs, we implemented an automatic tracking travel system in the intelligent wheelchair. 
The caregiver could simply operate the first wheelchair, and a second rear wheelchair automatically followed; hence, all the wheelchairs could be operated simultaneously.
In a nursing home care facility, many wheelchairs have to be moved at the same time when moving the care recipients from a room to a dining room or a bathroom. 
At such times, controlling a plurality of wheelchairs simultaneously could shorten travel time.

\subsubsection{Implementation}
We used two electric wheelchairs. 
We attached the tablet that displayed the artificial reality (AR) marker on the back of the base unit wheelchair. 
The slave unit wheelchair recognized the AR marker with its attached web camera. 
The slave unit followed the base unit wheelchair based on the position of the AR marker. 
The base unit could be remotely operated using HMD or by a controller standing beside a wheelchair. 

\begin{figure}[tb]
	\centering
    \includegraphics[width=\linewidth]{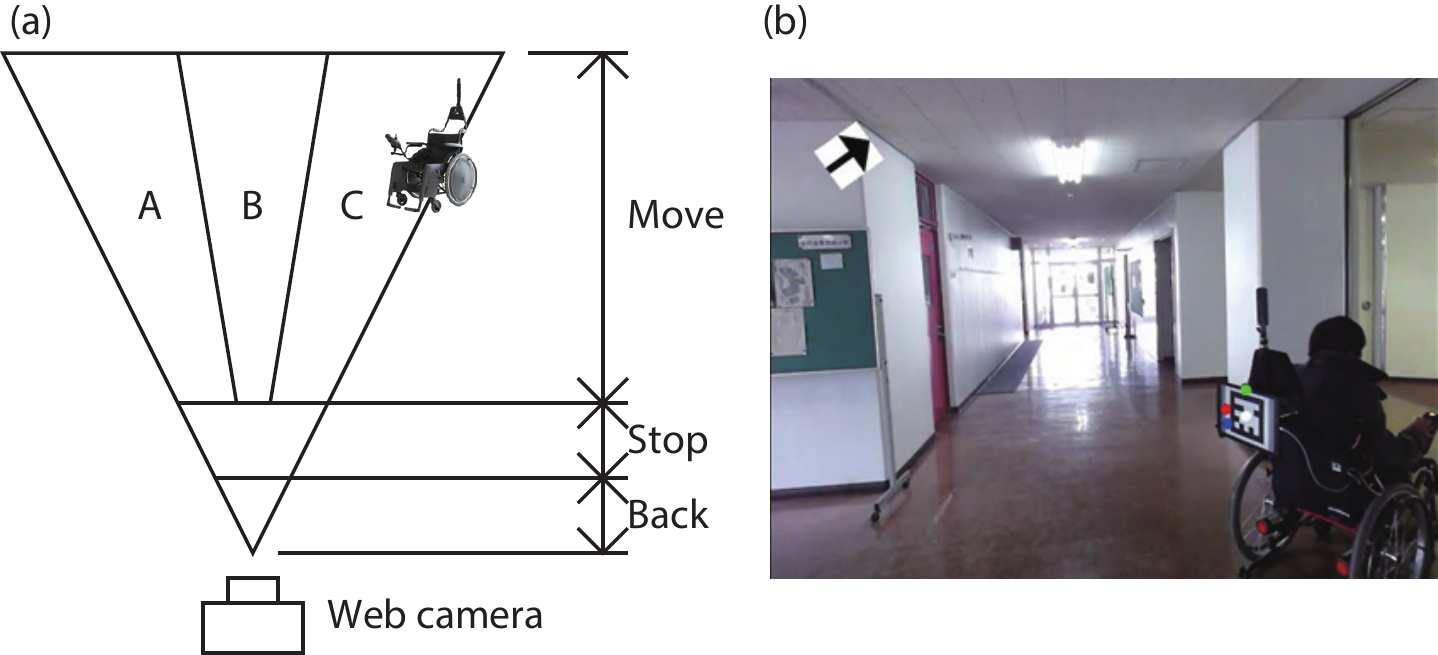}
    \caption{System of automatic tracking travel: (a) rear wheelchair utilizes the AR marker to acquire the position of the front wheelchair. By controlling the rear wheelchair according to the position of the AR marker, we succeeded in following using only the camera, and (b) image of the camera attached to the rear wheelchair.}
    \label{fig:system_traveling}
\end{figure}

We acquired three-dimensional (3D) position coordinates and roll, pitch, and yaw with AR markers using a processor running on a Windows laptop. 
The speed and direction of rotation of the slave unit wheelchair were calculated using the position of the AR marker. 
The speed of the slave unit was controlled according to the distance between the AR marker and the slave wheelchair. 
As shown in the Figure \ref{fig:system_traveling}(a), when approaching the base unit, the slave wheelchair stopped or backed up. 
The rotation direction of the slave unit was calculated according to the horizontal position of the base unit.
The image taken with the camera is divided into three areas as shown in Figure \ref{fig:system_traveling} (b). 
The rotation direction was determined by the area where the base unit wheelchair was located. 
The slave unit wheelchair turned right if the base unit wheelchair was in the right area, went straight if in the middle area, and turned left if in the left area. 
The rotation speed for turning right and left increased as the base unit wheelchair moved away from the center of the screen. 
In this system, we divided the image in the ratio 7 : 6 : 7, as left turn : straight ahead : left turn. 
When the slave unit rotated according to the rotation of the base unit, the slave unit started to turn as soon as the base unit started to turn. 
By using the proposed divide-by-three method, it was not necessary to record the locus of the base unit, and the slave unit could pass through the path that the base unit passed.

\subsubsection{Procedure}
We informed the participant about the study and obtained their consent as described above.
First, participants were asked to introduce themselves. 
Next, the experimenter introduced the functions of the intelligent wheelchair to the participants and explained the demonstration to be conducted later.
After the explanation was completed, we demonstrated the automatic tracking travel system. 
The demonstration time was 45 min. Finally, we asked the participants to provide their impressions of the demonstration.

\subsubsection{Participant:}
Two participants (two males), average age 22 years, participated in the experiment (Table \ref{tab:summary_driving_caregivers}). 
Both participants were caregivers. 
Both participants were employed in nursing care facilities that conducted workshops. One participant did not have prior knowledge of the Telewheelchair, and another participant knew a little. Both participants spoke Japanese.

\begin{table}[t]
\caption{Summary of workshop participants on follow-up driving for caregivers}
\vspace{1em}
 \begin{tabular}{l|c|c|c|c}
     & Sex & Age & Position & \small
     \begin{tabular}{l}
     Length of \\service (months) 
     \end{tabular}
     \\ \hline
    $P_{2}1$ & M & 22 & Caregiver & 3\\ \hline
    $P_{2}2$ & M & 21 & Caregiver & 21
  \end{tabular}
  \label{tab:summary_driving_caregivers}
\end{table}

\section{Findings of Study 1}
\subsection{Problem finding phase}

\subsubsection{Problems of care}
In this section, we present the problems at the nursing care facility that caregivers identified during the workshop.
First, the participants described the privacy of the older adults. 
$P_{1}2$ said, ``\textit{We need to give care in the toilet, but from the privacy point of view it is difficult to stay in the toilet.}'' 
$P_{1}4$ answered that taking care of meals is an example, ``\textit{Consideration must be given so that the people do not feel like they are being watched while I am observing the situation.}''

Next, participants described the time available for each older adult. 
$P_{1}2$ and $P_{1}5$ responded about the amount of work done by one caregiver, ``\textit{It is necessary for one person to change more than five diapers.}'' ($P_{1}2$), ``\textit{I do three person meals assistance at the same time.}'' ($P_{1}2$), and ``\textit{There was no time to communicate with the older adults because only two people assisted the bathing of 30 people a day.}''($P_{1}5$). 
$P_{1}4$ said about various assistance, ``\textit{the time spent per capita is short, because doing a lot of work within the day seems like non-stop work.}''

$P_{1}2$ and $P_{1}3$ felt that caregivers were always considering how to ensure older adults to live safely. 
$P_{1}2$ felt, ``\textit{Caregivers talk to the older adults sitting on a wheelchair from behind them}''. 
$P_{1}3$ said, ``\textit{Caregivers are always working hard because they care about multiple older adult people; but, it is necessary to have the older adults with the caregiver feel safe, so they should always respond slowly and clearly, with a smile.}'' 
$P_{1}2$ answered, ``\textit{Caregivers should always accompany the older adults for their safety.}''

\subsubsection{Problem solving methods}
We first describe the problem solving methods given by the participants before the technical learning seminar.
$P_{1}1$, $P_{1}2$ and $P_{1}6$ cited the problem of sitting when the older adults are in a wheelchair.
Because $P_{1}1$ wants to have the older adults sit comfortably in the wheelchair with little effort, he proposed a wheelchair whose seat surface could be adjusted. 
$P_{1}6$ replied, ``\textit{We will make the seat height changeable and make the cushioning material adjustable.}'' 
$P_{1}4$ and $P_{1}6$ proposed a solution for transferring. 
$P_{1}4$ replied, ``\textit{In order to enable transfer without relying on the caregiver's skills, we will make wheelchairs with sizes and functions suited to the older adults.}'' 
$P_{1}6$ said, ``\textit{I want the wheelchair's footrest and foot support to be able to close together automatically so as to solve the problem of moving the wheelchair from the bed.}'' 
$P_{1}1$ and $P_{1}5$ mentioned the time and labor of movement. 
$P_{1}1$ said that the trouble with moving the wheelchair can be solved if the wheelchair can be operated close to the older adults automatically. 
$P_{1}5$ replied, ``\textit{I hope all the assistance can be operated by a voice control system.}'' 
$P_{1}3$ answered, ``\textit{I want to walk in parallel rather than pushing the wheelchair from the back; I want the wheelchair to be operated remotely.}''

After the technical learning seminar, the problem-solving methods were discussed. 
$P_{1}1$ and $P_{1}2$ said that the older adults should be able to transfer alone and go to their destination. 
They said that it would be significantly effective to have automatic driving, voice recognition, wheelchair tracking, and calling for help functions on the wheelchair itself. 
$P_{1}5$ and $P_{1}2$ added a comment about dementia. 
$P_{1}5$ said, ``\textit{Older adults with dementia will forget their location and wheelchair operation procedures. 
For this kind of person to maneuver the wheelchair, it is necessary to control it by speech recognition. 
Furthermore, functions of remembering the route and driving automatically are desirable.}''

\subsubsection{Intelligent wheelchair demonstration}
We first performed the demonstration with the control operation of using a joystick mounted on an electric wheelchair. 
$P_{1}2$ felt, ``\textit{I could drive with my own will.}''
$P_{1}5$ answered, ``\textit{I can operate the driving just with my fingers, so people without paralysis in hand may also control it themselves.}'' $P_{1}2$, $P_{1}6$ replied, ``\textit{It is peace of mind.}''

Next, we performed the demonstration with the external wheelchair controller operated by an experimenter standing next to the participants. 
$P_{1}2$ and $P_{1}5$ answered, ``\textit{When seated in the wheelchair, I was relieved to see the face of the other person and to talk with him.}''  
$P_{1}3$ said, ``\textit{The operation became easy when I became familiar with the controller.}'' 
$P_{1}5$ answered, ``\textit{It is difficult to get used to the operations when I am using the controller.}'' 
$P_{2}3$ and $P_{2}4$ answered, ``\textit{The operation of the controller is difficult so sometimes I cannot control the wheelchair well.}''

Finally, the participants experienced a demonstration simulating automatic driving. 
$P_{1}2$ did not feel like a human. 
$P_{1}3$ replied, ``\textit{It is more comfortable than being operated using the controller.}'' 
$P_{1}5$ answered, ``\textit{It will be fun to move in this way because I can do other things while riding.}'' 
On the other hand, there were opinions such as ``\textit{I feel anxiety that there is no trust relationship with someone.}'' ($P_{1}4$)

\subsection{Problem solving phase}
Participants experienced a wheelchair equipped with cooperative operation functions. 
$P_{2}1$ felt that, ``\textit{It could be effective if used in nursing care facilities. 
In the cafeteria, I felt that this way could lead all of us to the dining room at almost the same time, without deciding the priority among the older adults. 
It may also change the living situation of people who cannot push wheelchairs by themselves. 
Those who live in nursing homes are often in their rooms, but I think that if they have this wheelchair they will be able to go shopping and go out at any time.}'' 
$P_{2}2$ mentioned that, ``\textit{It is natural to go straight, but I hope there will be a function that prevents it from hitting obstacles on the side or getting stuck even if it hits someting. 
For those with a paralyzed body, there is a possibility of falling, in that case I would like the wheelchair to stop urgently.}''

\section{Study 2: Interview with Older Adults}
\subsection{Goal}
The purpose of the experiment in this section was to investigate how older adult people accept intelligent electric wheelchairs. 
Artificial intelligence and automatic driving are relatively familiar to young people. 
However, they are not familiar to older adults; hence, it was unknown how the older adults feel about the new wheelchair. 
By experiencing the new electric wheelchair we developed, we can learn how older adults feel about technologies such as automatic driving and artificial intelligence. 
In addition, because previous experiments \cite{Hashizume:2018:TRC:3174910.3174914} were conducted on the same wheelchair from the viewpoint of the pilot, we can compare those results with the current experiment from the passenger's point of view.

\begin{figure*}[tb]
	\centering
    \includegraphics[width=\linewidth]{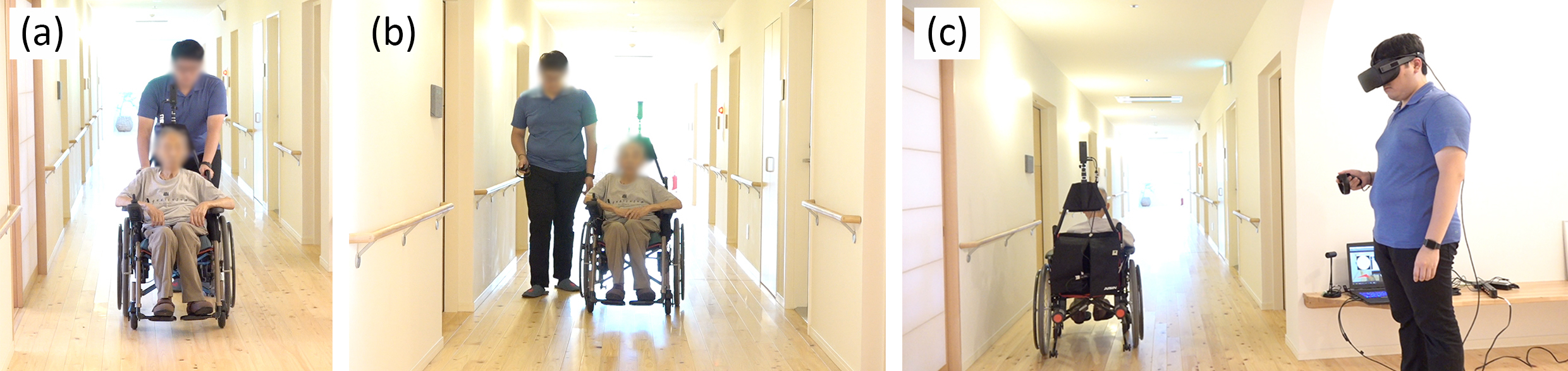}
    \caption{We carried out the following three pattern experiments with the participants on a wheelchair: (a) push the wheelchair from behind, (b) operate with the controller while walking or standing by the wheelchair, and (c) operate away from the wheelchair using HMD.}
    \label{fig:three_pattern}
\end{figure*}

\begin{figure}[tb]
	\centering
    \includegraphics[width=\linewidth]{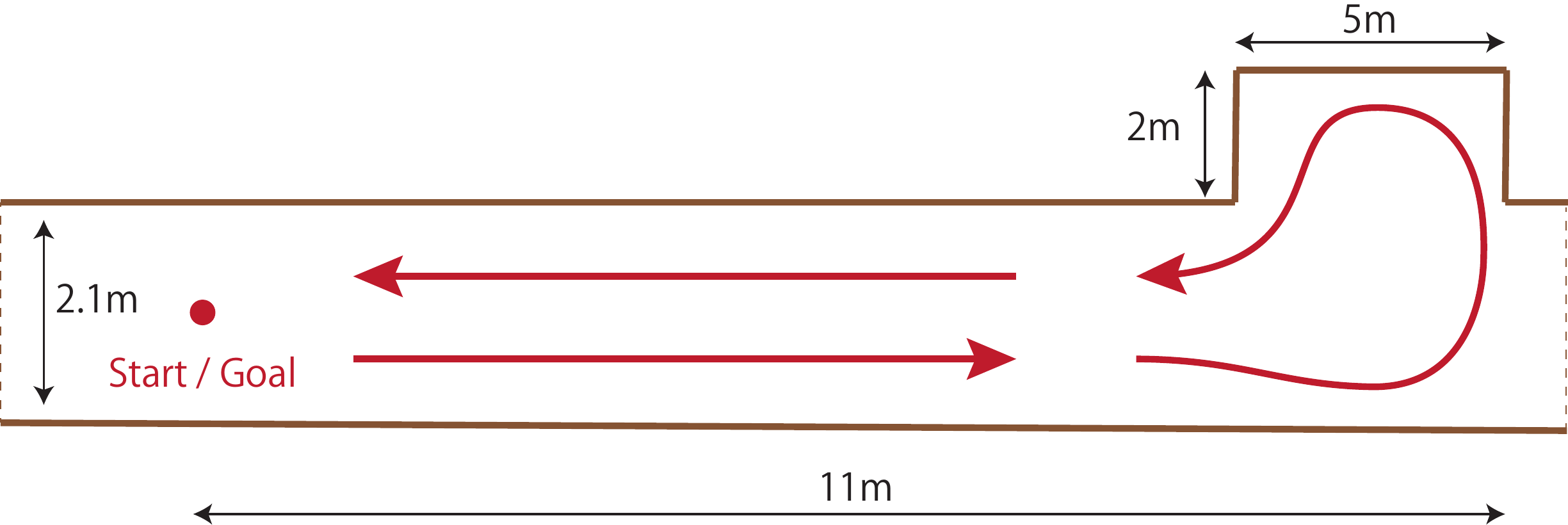}
    \caption{The course we used for the experiment at the nursing home. In the experiment we traveled back and forth 11 m in a corridor with a width of 2.1 m. The turning point is in the hall and the width of the corridor is widened.}
    \label{fig:cource}
\end{figure}

\subsection{Design}
In the previous study \cite{Hashizume:2018:TRC:3174910.3174914}, an experiment on the operability of the electric wheelchair was conducted. 
A remotely controllable electric wheelchair was developed. 
Four operation methods and  experiments on operability were conducted. 
We conducted experiments using three manipulation methods out of the four methods conducted in the previous research \cite{Hashizume:2018:TRC:3174910.3174914}: 
(i) Normal mode. 
This mode is a general operation method operated with the handle of the wheelchair (Figure \ref{fig:three_pattern}(a)); 
(ii) Stand by mode. 
Participants operate using the controller while standing next to the wheelchair (Figure \ref{fig:three_pattern}(b)); and (iii) HMD mode. 
Participants wear an HMD and operate the external controller. 
Participants can operate while freely watching the surroundings (Figure \ref{fig:three_pattern}(c)).
The older adults rode in the wheelchairs while the experimenter controlled it.
The wheelchair traveled straight through at corridor inside the nursing home for 11 m from the start point, performed a U-turn, and returned straight 11 m to the goal point (Figure \ref{fig:cource}). 
The width of the corridor was 2.1 m.

\subsection{Procedure}
We informed the participant about the study and obtained their consent as described above.
Participants experienced wheelchair operation in the order of Normal mode, Standby mode, and HMD mode. 
After experiencing each maneuvering mode, the participants were asked to answer the question about whether the ride was good or fearful, or whether they would want to use the wheelchair again. 
The answer to each question is based on a five-level Likert scale.

\subsection{Participants}
 Two participants (one female, one male), with an average age of 92.5 years, participated in the experiment (Table \ref{tab:summary_care_receivers}). 
Both of the participants lived in the nursing facility where the workshop was conducted. 
Participants required wheelchairs and could maneuver wheelchairs on their own if they were indoors. 
Participants could conduct commonplace conversations and write without problems. 
The wheelchair usually used is Next Core Puchi \footnote{\url{http://www.matsunaga-w.co.jp/search/detail_64.html}, (last accessed December 22, 2019. In Japanese)}.
All participants spoke Japanese.

\begin{table}
\caption{Summary of intelligent wheelchair demonstration participants for care receivers}
\vspace{1em}
\centering
 \begin{tabular}{l|c|c}
     & Sex & Age \\ \hline
    $P_{3}1$ & F & 92 \\ \hline
    $P_{3}2$ & M & 93
  \end{tabular}
  \label{tab:summary_care_receivers}
\end{table}

\section{Findings of Study 2}
First, the participants experienced the wheelchair that the experimenter pushed from behind. 
$P_{3}1$ answered, ``\textit{It's the easiest to get on
the wheelchair I have ever had.}'' 
$P_{3}2$ did not feel any change. 
Next, the participants got on a wheelchair operated by the external controller. 
$P_{3}1$ said, ``\textit{It was exhilarating, and I felt safe because there was the person next to me.}'' 
$P_{3}2$ answered, ``\textit{The wheelchair was slightly shaky.  There was no fear, but I think it is not much relative to the presence of a person next to me.}'' 
Finally, the older adults got on a remotely controlled wheelchair. 
$P_{3}1$ said, ``\textit{I think I'm a little afraid when there is no one next to me now, but if I get used to it I think that it is okay. I want to go outside but I was still too scared to do that.}'' 
$P_{3}2$ said, ``\textit{I was afraid of what will happen because the wheelchair is not under my control. But as the performance of the wheelchair rises, I think that fear may disappear.}'' 
$P_{3}1$ answered that the method of using the controller was the best among the three riding methods.

\section{Discussion}
\subsection{Busyness of caregivers}
In this section, we discuss the problems found during the practical workshop to find out how to make use of the intelligent electric wheelchair. 
In the opinion of the caregivers, the time to devote to nursing care for each older adults  is less because the actual nursing care work includes many aspects such as excretion aid, meal assistance, and bath assistance. 
Furthermore, in nursing care facilities, the number of older adults who need care from each caregiver is large because of caregiver shortage. 
As $P_{1}2$ and $P_{1}5$ stated, caregivers have shorter contact times with each older adults. 
Also mentioned from $P_{1}5$, it is difficult for a caregiver to communicate sufficiently with older adults. 
Moreover, if communication becomes impossible and it becomes like routine work, it may lead to at psychological burden on the older adult. 
In order to increase the time to be in contact with the older adults, it is necessary to reduce the time taken for nursing care.

Thus, we provided possible solutions for using an electric wheelchair to reduce person movement time. 
In nursing care facilities, there are many opportunities to move multiple wheelchairs at the same time from, for example, the living room to the dining room and the bathroom. 
If the caregiver can shorten the time it takes to move the wheelchairs, he or she can increase the time for other nursing care work. 
In order to shorten the time it takes to move, we installed a cooperative-operation function on an electric wheelchair. 
The function enables the system to move multiple wheelchairs at the same time, so the time it takes the caregiver to make the moves decreases.
We conducted a demonstration and interview with caregivers using an electric wheelchair equipped with the cooperative operation function. 
As a result of the interview, the participants recognized the usefulness of the cooperative-operation function. 
However, the caregivers believed it to be difficult for a single caregiver to monitor multiple wheelchairs simultaneously. 
Thus, a future improvement will be a monitoring system to prevent a passenger from falling down and accident.

\subsection{Operation Method}
The participants of the workshop provided many opinions on the operation method.
As $P_{1}5$ and $P_{1}2$ described, there are older adults with dementia living in nursing homes. 
Such older adults cannot remember how to operate devices, so it is difficult to use an electric wheelchair. 
In order to enable such older adults to use the electric wheelchair, it is necessary to be able to control the electric wheelchair without remembering the operation method themselves. 
A voice manipulation function was thought to be effective solution, because even older adults with dementia can control it thorough voice commands. 
In addition, older adults with dementia cannot remember the location of their room, so it is difficult for them to act alone.
An electric wheelchair that could automatically advance to the destination, by voice control and a navigation system, can reach the destination even for an older adults who has forgotten the location of the room.

From the workshop, changes to the operation method were suggested so that the older adults can be relieved of their fears. 
As mentioned from $P_{1}2$ and $P_{1}3$, it is important to make it possible for older adult people to feel secure. 
Also mentioned from $P_{1}2$, when a caregiver is assisting a wheelchair and pushing it from the back, the caregiver often talks to the older adult when behind their back.
From behind, the older adults cannot see the caregiver's facial expression, which makes it difficult to communicate. 
Therefore, this is possibility a situation that makes the older adults become uneasy. 
Hence, we implemented a function that permits the caregiver to stand next to the wheelchair and steer it. 
Standing next to the wheelchair makes it easier to communicate with the older adults. 
We implemented this control function using an external controller and demonstrated it to the older adults. 
After this demonstration, the older adults expressed that they felt safe because they could see the face of the operator.

\subsection{Psychology of the older adults}
In this part of the study, we conducted experiments with the older adults in the nursing home using the intellectualized electric wheelchair we developed previously. 
Overall, the older adults expressed a positive opinion. 
When the experimenter assisted by standing next to the wheelchair, there was a sense of safety, due to the presence of a person. 
In the case of remote control, the older adults felt fear. 
However, $P_{3}1$ thought that the fear may disappear if one became accustomed to the remote operation, and $P_{3}2$ said that the fear would go away if operational safety is higher. 
Therefore, we inferred that older adults might be more accepting of the remote control if safety could be ensured. 
It seemed that older adults positively understand the evolution of nursing care devices by technologies such as remote control. 
We need to experiment with the electric wheelchair equipped with automatic driving and artificial intelligence, to see if the older adults will accept it. 
However, we must always be conscious of the necessity of enhancing the safety of intelligent electric wheelchair.

\subsection{Limitations of our study}
In this section, we discuss the limitations of our experiments. 
First, we only held a one-day workshop and we have not conducted long-term experiments. 
Whether the intelligent electric wheelchair proposed by us can be used for nursing care work cannot be determined unless caregivers use it for a longer term. 
We have to conduct more experiments at a nursing facility for few month.

The participants did not have a wide range of symptoms. 
The older adults who participated in the experiments were able to talk and able to maneuver wheelchairs on their own. 
Furthermore, the number of subjects in some experiments was as few as two. 
There were some older adults living in the facility who had severe dementia and others who could not get up by themselves. 
For older adults with such severe disabilities, we need to experiment to see if intelligent electric wheelchairs are useful. 
For that reason, we need to design experimental methods for those who cannot talk well.

\section{Conclusion}
We conducted a workshop for caregivers and tried to identify some problems of nursing care and some solutions. 
There were some major problems described participants, however in the experiments we focused on the problem of a caregiver's workload. 
In order to solve this problem, we implemented cooperative-operation function to the wheelchair  and demonstrated it to caregivers. 
The usefulness of our proposed wheelchair was shown. 
In the interviews with older adults, we found that the older adults positively accepted the intelligent electric wheelchair.

\balance
\bibliographystyle{unsrt}
\bibliography{bibliography}

\end{document}